\documentclass[12pt] {revtex4}
\begin{document}
\title[Short Title]{Controlled Quantum Dense Coding in a Four-particle
Non-maximally Entangled State via Local Measurements}

\author{Chang-Bao \surname{FU}, Yan \surname{XIA}, Bo-Xue LIU, Shou \surname{ZHANG}\footnote{E-mail: szhang@ybu.edu.cn} }
\affiliation{Department of Physics, College of Science and
Engineering, Yanbian University, Yanji, 133002, PR China}

\author{Kyu-Hwang \surname{YEON}}
\affiliation{Department of Physics,  Institute for Basic  Science
Research, College of Natural Science, Chungbuk National University,
Cheonju, Chungbuk 361-763}

\author{Chung-In \surname{UM}}
\affiliation{Department of Physics, College of Science, Korea
University, Seoul 136-701}

\begin{abstract}
A controlled quantum dense coding scheme is investigated with a
four-particle non-maximal quantum channel. \ The amount of
classical information is shown to be capable of being controlled
by the controllers through adjustments of the local measurement
angles and to depend on the coefficients of the quantum channel; \
in addition, the four particles are distributed in two inverse
ways in such an quantum channel. \ A restricted condition for
distributing the particles to realize quantum
dense coding in an arbitrary ($N+2$)-particle quantum channel is proposed. 
{\bf PACS} number(s): {\it 03.67.Hk, 03.65. Ud, 03.67. -a}
\end{abstract}
\maketitle Quantum entanglement plays a key role in quantum
information theory and teleportation \cite{0001,0002}. \ Quantum
dense coding (QDC) \cite{0003,0004,0005,0006} is one of the
exhibitions of entanglement in quantum communication. \ Normally,
the classical capacity of a transmission channel is 1 bit;\ \
however, in dense coding, with the help of entanglement, people
can transmit two bits of classical information by sending only one
qubit. \ Bose {\it et al.} and Lee {\it et al.} \cite{0007,0008}
and Bose \cite{0009} have generalized QDC between two parties to
multiparties and mixed state dense coding, respectively.

On the other hand, Hao {\it et al.} \cite{0010} have proposed a
controlled dense coding scheme by using the three-particle
Greenberger-Horne-Zeilinger (GHZ) state. \ In this scheme, one
party (Alice) can transmit information to the second party (Bob)
whereas the local measurement of the third party (Cliff) serves as
quantum erasure. \ Cliff can control the quantum channel between
Alice and Bob via a local measurement to realize controlled dense
coding between Alice and Bob. \ Chen and Kuang \cite{0011} have
generalized the controlled dense coding scheme of the
three-particle GHZ quantum channel to the case of an (N +
2)-particle GHZ quantum channel via a series of local
measurements.

In this paper, we study controlled quantum dense coding in a
four-particle non-maximal quantum channel via local measurements.
\ Our goal consists of three aspects: \ (i) Study how the
transmitted amount of classical information is controlled by the
controllers through adjustments of the local measurement angles
and how it depends on the coefficients of the entangled quantum
channel. \ (ii) Discuss the distribution of the four particles in
such a four-particle non-maximal quantum channel. \ (iii) Propose
a restricted condition for how to distribute the particles to
realize quantum dense coding in an arbitrary ($N+2$)-particle
quantum channel.

Firstly, we review the QDC scheme. \ Let us assume that Alice and
Bob initially share the Bell state $|\phi\rangle^{+}$. \ Locally
operating on her qubit, Alice obtains the four orthogonal Bell
states $\hat{I}|\phi\rangle^{+}=|\phi\rangle^{+}$,
$\hat{\sigma}_{xA}|\phi\rangle^{+}=|\psi\rangle^{+}$,
$\hat{\sigma}_{yA}|\phi\rangle^{+}=i |\psi\rangle^{-}$, and
$\hat{\sigma}_{zA}|\phi\rangle^{+}=|\phi\rangle^{-}$. \ Alice then
sends her qubit to Bob. By making a Bell measurement, Bob is able
to obtain two bits of classical information. \ The four Bell
states are defined by
\begin{equation}\label{1}
|\phi^{\pm}\rangle=\frac{1}{\sqrt{2}}\ (|00\rangle\pm|11\rangle),
\end{equation}
\begin{equation}\label{2}
|\psi^{\pm}\rangle=\frac{1}{\sqrt{2}}\ (|01\rangle\pm|10\rangle).
\end{equation}

Secondly, we now propose our scheme. \ Alice (party 2) and Bob
(party 3) share a four-particle non-maximal quantum channel
\begin{equation}\label{3}
|\psi\rangle=(a|0000\rangle+b|1001\rangle+c|0110\rangle+d|1111\rangle)_{1234},
\end{equation}
where the coefficients $a$, $b$, $c$, and $d$ are real, and
$|a|^{2} + |b|^{2} + |c|^{2} + |d|^{2}$ = 1. \ We assume that
qubit 2, qubit 3, qubit 1, and qubit 4 belong to Alice (party 2),
Bob (party 3), party 1, and party 4, respectively.

We suppose that party 4 carries out a unitary operation on his
qubit 4 in the following forms:
\begin{equation}\label{4}
|+\rangle_{4}=\cos\theta_{1}|0\rangle_{4}+\sin\theta_{1}|1\rangle_{4},
\end{equation}
\begin{equation}\label{5}
|-\rangle_{4}=\sin\theta_{1}|0\rangle_{4}-\cos\theta_{1}|1\rangle_{4};
\end{equation}
then, the four-particle non-maximal quantum channel can be
rewritten as
\begin{equation}\label{6}
|\psi\rangle=|\varphi\rangle_{123}\otimes|+\rangle_{4}+|\phi\rangle_{123}\otimes|-\rangle_{4},
\end{equation}
where
\begin{equation}\label{7}
|\varphi\rangle_{123}=(a \cos\theta_{1}|000\rangle+b
\sin\theta_{1}|100\rangle+c \cos\theta_{1}|011\rangle+d
\sin\theta_{1}|111\rangle)_{123},
\end{equation}
\begin{equation}\label{8}
|\phi\rangle_{123}=(a \sin\theta_{1}|000\rangle-b
\cos\theta_{1}|100\rangle+c \sin\theta_{1}|011\rangle-d
\cos\theta_{1}|111\rangle)_{123}.
\end{equation}
Party 4 can obtain two probable local measurement results from
Eq.~(\ref{6}). \ One is $|+\rangle_{4}$, for which the state of
qubits 1, 2, 3 collapses to $|\varphi\rangle_{123}$; \ the other
is $|-\rangle_{4}$, for which the state of qubits 1, 2, 3
collapses to $|\phi\rangle_{123}$. \ We only consider the case in
which the fourth party's measurement result is $|+\rangle_{4}$,
for which the state of qubits 1, 2, 3 collapses to
$|\varphi\rangle_{123}$ in Eq.~(\ref{7}).

Then, party 1 carries out a unitary operation on his qubit 1 in
the following forms:
\begin{equation}\label{9}
|+\rangle_{1}=\cos\theta_{2}|0\rangle_{1}+\sin\theta_{2}|1\rangle_{1},
\end{equation}
\begin{equation}\label{10}
|-\rangle_{1}=\sin\theta_{2}|0\rangle_{1}-\cos\theta_{2}|1\rangle_{1}.
\end{equation}
The three-particle state $|\varphi\rangle_{123}$ can be rewritten
as
\begin{equation}\label{11}
|\varphi\rangle_{123}=|\varphi\rangle_{23}\otimes|+\rangle_{1}+|\phi\rangle_{23}\otimes|-\rangle_{1},
\end{equation}
where
\begin{equation}\label{12} |\varphi\rangle_{23}=(a
\cos\theta_{1} \cos\theta_{2}+b \sin\theta_{1}
\sin\theta_{2})|00\rangle_{23}+(c \cos\theta_{1} \cos\theta_{2}+d
\sin\theta_{1} \sin\theta_{2})|11\rangle_{23},
\end{equation}
\begin{equation}\label{13}
|\phi\rangle_{23}=(a \cos\theta_{1} \sin\theta_{2}-b
\sin\theta_{1} \cos\theta_{2})|00\rangle_{23}+(c \cos\theta_{1}
\sin\theta_{2}-d \sin\theta_{1} \cos\theta_{2})|11\rangle_{23}.
\end{equation}
Party 1 can also obtain two probable local measurement results
from Eq.~(\ref{11}). \ If the measurement result is
$|+\rangle_{1}$, the state of qubits 2, 3 collapses to
$|\varphi\rangle_{23}$. \ Otherwise, the state of qubits 2, 3
collapses to $|\phi\rangle_{23}$. \ We only consider the case in
which the first party's measurement result is $|+\rangle_{1}$; \
then, the state of qubits 2, 3 collapses to $|\varphi\rangle_{23}$
in Eq.~(\ref{12}).

After receiving the measurement results and information on the
measurement angles from party 4 and party 1, Alice and Bob can
obtain the two-particle maximally entangled state from the
two-particle non-maximally entangled state in Eq.~(\ref{12}). \ If
Alice induces an auxiliary qubit $|0\rangle_{\rm a}$ and performs
the unitary operation
\begin{equation}\label{14}
\hat{U}_{2\ \rm a}=\left(%
\begin{array}{cccc}
\tan\gamma & 0 & \sqrt{1-\tan^{2}\gamma} & 0 \\
0 & 1 & 0 & 0 \\
\sqrt{1-\tan^{2}\gamma} & 0 & -\tan\gamma & 0 \\
0 & 0 & 0 & -1 \\
\end{array}%
\right)
\end{equation}
on her qubit 2 and on the auxiliary qubit, which are written under
the basis \{$|0\rangle_{2}$$|0\rangle_{\rm a}$,
$|1\rangle_{2}$$|0\rangle_{\rm a}$, $|0\rangle_{2}$$|1\rangle_{\rm
a}$, $|1\rangle_{2}$$|1\rangle_{\rm a}$\}. \ In the unitary
transformation of Eq.~(\ref{14}), $\tan\gamma$ is expressed by
\begin{equation}\label{15}
\tan\gamma=\frac{c\ \cos\theta_{1}\ \cos\theta_{2}+d\
\sin\theta_{1}\ \sin\theta_{2}}{a\ \cos\theta_{1}\
\cos\theta_{2}+b\ \sin\theta_{1}\ \sin\theta_{2}}.
\end{equation}
The collective unitary operation $\hat{U}_{2\ \rm a}$$\otimes$\
$\hat{I}_{3}$ transforms the direct product state
$|\varphi\rangle_{23}$\ $\otimes$\ $|0\rangle_{\rm a}$ to a
three-particle entangled state:
\begin{eqnarray}\label{16}
|\varphi\rangle_{23\rm a}&=&\sqrt{2}\ \sin\gamma\
 |\phi^{+}\rangle_{23}\otimes|0\rangle_{\rm a}+\cos\gamma\ \sqrt{1-\tan^{2}\gamma}\ |00\rangle_{23}\otimes|1\rangle_{\rm a},
\end{eqnarray}
where $|\phi^{+}\rangle_{23}$ is one of the Bell states of qubit 2
and qubit 3 as given by Eq.~(\ref{1}), $|00\rangle_{23}$ is the
unentangled state of the two qubits, and the parameter angle
$\gamma$ is defined by
\begin{equation}\label{17}
\sin\gamma=\frac{1}{\sqrt{e}}\ (c\ \cos\theta_{1}\
\cos\theta_{2}+d\ \sin\theta_{1}\ \sin\theta_{2}),
\end{equation}
\begin{equation}\label{18}
\cos\gamma=\frac{1}{\sqrt{e}}\ (a\ \cos\theta_{1}\
\cos\theta_{2}+b\ \sin\theta_{1}\ \sin\theta_{2}),
\end{equation}
with
\begin{equation}\label{19}
e=(a\ \cos\theta_{1}\ \cos\theta_{2}+b\ \sin\theta_{1}\
\sin\theta_{2})^{2}+(c\ \cos\theta_{1}\ \cos\theta_{2}+d\
\sin\theta_{1}\ \sin\theta_{2})^{2}.
\end{equation}

Alice and Bob can obtain a two-particle maximally entangled state
when Alice measures the auxiliary qubit and obtains
$|0\rangle_{\rm a}$ from Eq.~(\ref{16}). \ From the above
procedure, we obviously see that party 4 and party 1 control the
entanglement between particle 2 and particle 3 with local
measurements. \ The average classical amount of information
transmitted from Alice to Bob adds up to
\begin{equation}\label{20}
C=1+2\ |\sin\gamma|^{2}=1 + 2\ [1+\cot^{2}\gamma]^{-1}.
\end{equation}

Thirdly, in order to expatiate on how the local measurement angles
from party 1 and party 4 affect the amount of information, we
discuss the expression of the transmitted classical amount of
information in Eq.~(\ref{20}). \ (i) For the case of $|\tan\gamma|
< 1$, the classical amount of information transmitted from Alice
to Bob is less than two bits from Eq.~(\ref{20}) with
Eq.~(\ref{15}). \ (ii) For the case of $\pi/4$ transformations and
$|a|$ = $|b|$ = $|c|$ = $|d|$ = $1/2$, party 1 and party 4 carry
out $\theta_{\rm 1}=\pi/4$ and $\theta_{\rm 4}=\pi/4$, and the
four-particle non-maximal quantum channel in Eq.~(\ref{3}) become
a maximally entangled state. \ From Eq.~(\ref{20}) with
Eqs.~(\ref{17})$-$(\ref{19}), we can see that the classical amount
of information transmitted from Alice to Bob reaches a maximal
value of two bits. \ Thus, we can conclude that the transmitted
classical amount of information not only depends on the
measurement angles, which are controlled by party 1 and party 2,
but also depends on the coefficients of the four-particle
non-maximal quantum channel.

Fourthly, we discuss the distribution of the four particles in
such a four-particle non-maximal quantum channel. \ From
Eq.~(\ref{3}) with Eq.~(\ref{1}) and Eq.~(\ref{2}), if quantum
dense coding is to be realized, the four particles must be
distributed in the following forms: \ (i) Particle 2 and particle
3 belong to Alice and Bob, respectively, and particle 1 and
particle 4 as quantum erasure; \ then, Alice and Bob can obtain a
two-particle non-maximally entangled state that is a linear
combinations of states $\{|00\rangle_{23}, |11\rangle_{23} \}$. \
We have described such the case in much greater detail in this
paper. \ (ii) If the particles are distributed in an opposite way,
Alice and Bob can again obtain a two-particle non-maximally
entangled state that is a linear combinations of states
$\{|00\rangle_{23}, |11\rangle_{23} \}$. \ The remaining
distribution of the four particles are unsuccessful. \ In a word,
there are only two ways to distribute four particles to realize
quantum dense coding in such a four-particle non-maximal quantum
channel.

Finally, we propose a restricted condition on how to distribute
the particles to realize quantum dense coding in an arbitrary
($N+2$)-particle quantum channel. \ Here, $N+2$ parties,
possessing one particle each, share an arbitrary ($N+2$)-particle
quantum channel. \ After $N$ particles are served as quantum
erasure via a series of local measurements, to realize controlled
quantum coding, the sender and the receiver must obtain a
two-particle non-maximally entangled state that must be a linear
combinations of states similar to one of states that are generated
after the sender encodes her qubit.

It must be stressed that our scheme is valuable. \ Controlled
quantum dense coding has been studied by others by employing a
maximally entangled state, but depending on the physical systems,
a maximally entangled state can't always be generated. \ We employ
a partially entangled state instead of a maximally entangled
state, which is convenient for physical systems. \ Our scheme in a
four-particle non-maximal quantum channel is an extension of the
controlled quantum dense coding scheme to other schemes employing
a GHZ state as quantum channel. \ We first propose a restricted
condition on how to distribute the particles to realize quantum
dense coding in an arbitrary ($N+2$)-particle quantum channel.

In summary, we have studied the QDC scheme between two fixed
particles in a four-particle non-maximal quantum channel. \ We
have found that the transmitted classical amount of information
can be controlled by the controllers through adjusting the local
measurement angles and that depends on the coefficients of the
four-particle non-maximal quantum channel. \ We have shown that
there are only two ways of distributing four particles in such a
four-particle non-maximal quantum channel to realize quantum dense
coding. \ We have proposed a restricted condition on how to
distribute the particles to realize quantum dense coding in an
arbitrary ($N+2$)-particle quantum channel.

\begin{center}
{\bf ACKNOWLEDGMENTS}
\end{center}

This work was supported by the Korea Science and Engineering
Foundation and by the National Natural Science Foundation of China
under Grant No. 60261002.

\end{document}